\documentclass[12pt]{article}

\usepackage[dvips]{graphicx}
\usepackage{epsfig}
\usepackage{bbm}
\usepackage{amsmath,amssymb,amstext}
\usepackage{array}
\usepackage{rotating}
\usepackage{a4}
\usepackage{a4wide}
\usepackage{color}
\usepackage[colorlinks=true,linkcolor=red,citecolor=green, filecolor=magenta,urlcolor=cyan,bookmarks=true,bookmarksopen=true]{hyperref}

\usepackage[all]{hypcap}
\usepackage{mciteplus}

\def\be{\begin{equation}}
\def\ee{\end{equation}}
\newcommand{\ba}{\begin{array}{c}}
\newcommand{\baz}{\begin{array}{cc}}
\newcommand{\bad}{\begin{array}{ccc}}
\newcommand{\bav}{\begin{array}{cccc}}
\newcommand{\baf}{\begin{array}{ccccc}}
\newcommand{\bena}{\begin{eqnarray}}
\newcommand{\eena}{\end{eqnarray}}
\newcommand{\bea}{\begin{equation} \begin{array}{c}}
\newcommand{\eea}{ \end{array} \end{equation}}
\newcommand{\ea}{\end{array}}

\begin{document}

\title{
\hfill {\small DFTT 55/2009}\\
\vglue -0.3cm
\hfill {\small SNUPT 09-013} \vskip 0.5cm
\Large \bf
{\color{red} Investigating light neutralinos at neutrino telescopes} }

\author{
V. Niro$^a$\thanks{email: \tt viviana.niro@mpi-hd.mpg.de}~~,
A. Bottino$^b$\thanks{email: \tt bottino@to.infn.it}~~,
N. Fornengo$^b$\thanks{email: \tt fornengo@to.infn.it}~~,
and S. Scopel$^c$\thanks{email: \tt scopel@phya.snu.ac.kr}
\\ 
{\normalsize \it$^a$Max--Planck--Institut f\"ur Kernphysik,}\\
{\normalsize \it  Postfach 103980, D--69029 Heidelberg, Germany} 
\\ 
{\normalsize \it$^b$Dipartimento di Fisica Teorica, Universit\`a di Torino}\\
{\normalsize \it and INFN, Sez. di Torino, via P. Giuria 1, I-10125 Torino, Italia}
\\ 
{\normalsize \it$^c$School of Physics and Astronomy, Seoul National University}\\
{\normalsize \it Gwanak-ro 599, Gwangak-gu, Seoul 151-749, Korea}}
\date{ \today}
\maketitle
\thispagestyle{empty}
\vspace{-0.8cm}
\begin{abstract}
\noindent {\color{blue}
We present a detailed analysis of the neutrino-induced muon signals
coming from neutralino pair-annihilations inside the Sun and the
Earth with particular emphasis for light neutralinos.
The theoretical model considered is an effective MSSM without
gaugino-mass unification, which allows neutralinos of light masses
(below 50 GeV). The muon events are divided in through-going and stopping
muons, using the geometry of the Super-Kamiokande detector. In the
evaluation of the signals, we take into account the relevant hadronic
and astrophysics uncertainties and include neutrino oscillation and
propagation properties in a consistent way.
We derive the ranges of neutralino masses which could be explored at
neutrino telescopes with a low muon-energy threshold (around 1 GeV)
depending on the category of events and on the values of the various
astrophysics and particle-physics parameters.
A final analysis is focussed to the upward muon fluxes which could be
generated by those neutralino configurations which are able to explain
the annual modulation data of the DAMA/LIBRA experiment. We show
how combining these data with measurements at neutrino telescopes could help
in pinning down the features of the DM particle and in
restraining the ranges of the many quantities (of astrophysics and particle-physics
origins) which enter in the evaluations and still suffer from large uncertainties. }

\end{abstract}

\newpage

\section{\label{sec:intro}Introduction}

In the papers of 
Ref.\,\cite{Bottino:2002ry,*Bottino:2003iu} it was shown that light
neutralinos with a mass in the range 7 GeV $\lesssim m_{\chi}
\lesssim$ 50 GeV are interesting candidates for particle dark matter,
with direct detection rates accessible to experiments with large
exposure (of order 100 000 kg day) and low energy threshold (of a few
keV). This population of light neutralinos arises in the Minimal
Supersymmetric extension of the Standard Model (MSSM) when the unification 
of gaugino masses at a Grand Unified (GUT) scale is not
assumed \cite{dlp:08}. 
Indeed, in this supersymmetric framework the lower bound of
$\sim$ 7 GeV on the neutralino mass is set by a cosmological bound on
the neutralino relic density
\cite{Bottino:2002ry,*Bottino:2003iu}. This is at variance with the
lower bound $m_{\chi} \gtrsim$ 50 GeV, which is derived from the LEP
lower limit on the chargino mass within the MSSM with gaugino mass
unification at the GUT scale.

A direct comparison of the theoretical predictions of
Ref.\,\cite{Bottino:2002ry,*Bottino:2003iu} with experimental data was
made possible when the DAMA Collaboration published the results of its
measurements collected with a NaI detector of 100 kg over 7 annual
cycles \cite{Bernabei:2003za}. Actually, in Ref.\,\cite{Bottino:2003cz}
it was proved that the population of light neutralinos fitted well
these DAMA results.

The inclusion of the channeling effect in the experimental analysis
\cite{Bernabei:2007hw} and the further results of the DAMA/LIBRA
combined data \cite{Bernabei:2008yi} confirmed the good agreement 
of the predictions of Ref.\,\cite{Bottino:2002ry,*Bottino:2003iu} 
with the experimental data, as discussed in 
Refs.\,\cite{Bottino:2007qg,Bottino:2008mf}.

Possible indirect effects of light neutralinos were studied in
Refs.\,\cite{Bottino:2008mf,Bottino:2004qi}, where it was shown that
present cosmic antiproton data set sizable constraints on the
supersymmetric space; measurements of cosmic antideuterons
\cite{Donato:1999gy,*Donato:2008yx,*Baer:2005tw} with forthcoming
airborne experiments \cite{Hailey:2005yx,*Ktalk:2007,*ams:2007} were
indicated in \cite{Bottino:2008mf,Bottino:2004qi} as a very promising
investigation mean for the light neutralino population.

In Ref.\,\cite{Bottino:2004qi} also possible signals induced in
neutrino telescopes by the neutrinos produced by pair-annihilations of
light neutralinos captured in the Earth and the Sun were
discussed. The analysis was carried out in the case of a single
category of events, those due to upward through-going muons.

In the present paper we resume the analysis of
Ref.\,\cite{Bottino:2004qi}, by implementing and extending it in
various distinctive features: a) the particle-physics uncertainties in
hadronic quantities which affect the capture rate of relic neutralinos
by the celestial bodies are taken into account and discussed in
detail; b) all main processes that occur during the neutrino
propagation:  neutrino oscillations and neutrino incoherent
interactions with matter are included; c) the investigation comprises
also the category of upward stopping muons (actually, this turns out
to be the most promising case for upward-going muons induced by pair
annihilation of light neutralinos).

Indirect evidence of WIMPs in our halo by measurements of upward-going 
muons at neutrino telescopes, generated by WIMP pair-annihilation in 
celestial bodies, has been the subject of many investigations 
in the past, see for instance 
\cite{Gaisser:1986ha,Drees:1993bh,
Silk:1985ax,*Freese:1985qw,*Giudice:1988vs,*Gelmini:1990je,
*Kamionkowski:1991nj,*Halzen:1991kh,*Mori:1993tj,
*Gandhi:1993ce,*Bergstrom:1996kp,*Bergstrom:1997tp,
Bottino:1991dy,*Bottino:1994xp,*Berezinsky:1996ga,*Bottino:1998vw,
*Bottino:2000gc}.

Recently, a number of papers appeared where possible signals at
neutrino telescopes due to pair annihilation of light DM particles 
are discussed. These consider either generic WIMPs
with assumed dominance of specific annihilation channels
\cite{Hooper:2008cf,Feng:2008qn} or discuss specific DM candidates other than
neutralinos: WIMPless dark matter and mirror dark matter
\cite{Feng:2008qn}, leptonically interacting dark matter
\cite{Kopp:2009et} and DM particles which directly annihilate to
neutrinos \cite{Kumar:2009ws}. All these investigations are restricted
to signals from the Sun.

At variance with these analyses, the present paper deals with neutrino
signals produced in cosmic macroscopic bodies by pair annihilation of
light neutralinos whose specific supersymmetric properties are those
which satisfy all existing particle-physics and cosmological
constraints, as analyzed in
Refs.\,\cite{Bottino:2002ry,*Bottino:2003iu,Bottino:2008mf}. Furthermore,
our investigation entails neutrino signals expected from the Sun as
well as those from the Earth.

Our results are given first for the whole population of light
neutralinos, then for those subsets which are selected by the
DAMA/LIBRA results \cite{Bernabei:2008yi}, when these are interpreted
in terms of relic neutralinos. We separate the case where the
channeling effect is included from the one where this effect is
neglected.

The scheme of the present paper is the following. In
Sect.\,\ref{sec:model} the main features of the supersymmetric model
employed here are summarized.  The formulae providing the capture
rates of relic neutralinos by celestial bodies and the annihilation
rates due to neutralino pair-annihilation are recalled in 
Sect.\,\ref{sec:capture}, together with some properties of the
hadronic quantities which enter in the evaluations of the
neutralino-nucleon cross section. The generation and the propagation 
of the neutrino fluxes is described in Sect.\,\ref{sec:neutrino}, 
their ensuing muon fluxes are derived in Sect.\,\ref{sec:muon}. 
Results and conclusions are given in Sect.\,\ref{sec:results}.

\section{\label{sec:model}Theoretical model}

The supersymmetric scheme we employ in the present paper is the one
described in
Refs.\,\cite{Bottino:2002ry,*Bottino:2003iu,Bottino:2008mf}: an
effective MSSM scheme (effMSSM) at the electroweak scale, with the
following independent parameters: $M_1, M_2, \mu, \tan\beta, m_A,
m_{\tilde q}, m_{\tilde l}$ and $A$. Notations are as follows: $M_1$
and $M_2$ are the U(1) and SU(2) gaugino masses (these parameters are
taken here to be positive), $\mu$ is the Higgs mixing mass parameter,
$\tan\beta$ the ratio of the two Higgs v.e.v.'s, $m_A$ the mass of the
CP-odd neutral Higgs boson, $m_{\tilde q}$ is a squark soft--mass
common to all squarks, $m_{\tilde l}$ is a slepton soft--mass common
to all sleptons, and $A$ is a common dimensionless trilinear parameter
for the third family, $A_{\tilde b} = A_{\tilde t} \equiv A m_{\tilde
q}$ and $A_{\tilde \tau} \equiv A m_{\tilde l}$ (the trilinear
parameters for the other families being set equal to zero). In our
model, no gaugino--mass unification at a Grand Unified (GUT) scale is
assumed. The lightest neutralino is required to be the lightest
supersymmetric particle and stable (because of R--parity
conservation).

The numerical analysis presented in the present paper was performed
by a scanning of the supersymmetric parameter space, with the
following ranges of the MSSM parameters: $1 \leq \tan \beta \leq 50$,
$100 \, {\rm GeV} \leq |\mu| \leq 1000 \, {\rm GeV}, 5 \, {\rm GeV}
\leq M_1 \leq 500 \, {\rm GeV}, 100 \, {\rm GeV} \leq M_2 \leq 1000 \,
{\rm GeV}, 100 \, {\rm GeV} \leq m_{\tilde q}, m_{\tilde l} \leq 3000
\, {\rm GeV }$, $90\, {\rm GeV }\leq m_A \leq 1000 \, {\rm GeV }$, $-3
\leq A \leq 3$.

The supersymmetric parameter space has been subjected to all available 
constraints due to accelerator data on supersymmetric and Higgs boson searches
(CERN $e^+ e^-$ collider LEP2 \cite{alephdelphi:01,*lep:2001xwa,*lep2:04} 
and Collider Detectors D0 and CDF at Fermilab 
\cite{Affolder:2000rg,*Abazov:2006fe}) and to other particle-physics 
precision results: measurements of the 
$b \rightarrow s + \gamma$ decay process \cite{Barberio:2006bi}
[2.89 $\leq B(b \rightarrow s + \gamma) \cdot 10^{-4} \leq$ 4.21 is employed
here], the upper bound on the branching ratio $BR(B_s^{0} \rightarrow 
\mu^{-} + \mu^{+})$ \cite{Abazov:2007iy} [we take $BR(B_s^{0} \rightarrow \mu^{-} 
+ \mu^{+}) < 1.2 \cdot 10^{-7}$] and measurements of the muon anomalous 
magnetic moment $a_\mu \equiv (g_{\mu} -2)/2$: for the deviation 
$\Delta a_{\mu}$ of the  experimental world average from the 
theoretical evaluation within the Standard Model we use here the range
$-98 \leq \Delta a_{\mu} \cdot 10^{11} \leq 565 $. For other details 
concerning these constraints see 
Refs.\,\cite{Bottino:2002ry,*Bottino:2003iu,Bottino:2008mf}.

Also included is the cosmological constraint that the neutralino relic
abundance does not exceed the maximal allowed value for cold dark
matter, {\it i.e.} $\Omega_{\chi} h^{2} \leq (\Omega_{CDM} h^{2})_{\rm
max}$.  We set $(\Omega_{CDM} h^{2})_{\rm max} = 0.122$, as derived at the 
2$\sigma$ level from the results of Ref.\,\cite{Dunkley:2008ie}. We
recall that this cosmological upper bound implies on the neutralino
mass the lower limit $m_{\chi} \gtrsim$ 7 GeV
\cite{Bottino:2002ry,*Bottino:2003iu}.

\section{\label{sec:capture}Capture rates and annihilation rates}

In the present section we summarize the main formulae which we employed to 
evaluate the capture rates of neutralinos by celestial
bodies and their pair-annihilation in neutrinos.

The capture rate $C$ of the relic neutralinos by a macroscopic body is
given by the standard formula \cite{Gould:1987ir,*Gould:1987ww,*Gould:1991ApJ}
\be
C=\frac{\rho_{\chi}}{v_{\chi}}\sum_{i}\frac{\sigma_{{\rm el},
i}}{m_{\chi}m_{i}}\, (M_{B}f_{i})\,\langle v^{2}_{esc}\rangle_i \,
X_i\,,
\label{eq:Capture}
\ee

\noindent where $v_{\chi}$ is the neutralino mean velocity, $\sigma_{{\rm el}, i}$ 
is the cross section of the neutralino elastic
scattering off the nucleus $i$ of mass $m_{i}$, $M_{B}f_{i}$ is the
total mass of the element $i$ in the body of mass $M_{B}$, $\langle
v^{2}_{esc}\rangle_{i}$ is the square escape velocity averaged over
the distribution of the element $i$, $X_{i}$ is a factor which takes
account of kinematical properties occurring in the neutralino--nucleus
interactions.

The neutralino local density is denoted by $\rho_\chi$. One can assume that
$\rho_{\chi}$ is equal to the local value of the total DM density
$\rho_0$, when the neutralino relic abundance $(\Omega_\chi h^{2})$
turns out to be at the level of a minimal $(\Omega_{CDM} h^{2})_{\rm min}$ consistent
with $\rho_0$. On the contrary, when $(\Omega_\chi h^{2})$ is smaller
than $(\Omega_{CDM} h^{2})_{\rm min}$, the value to be assigned to
$\rho_{\chi}$ has to be appropriately reduced.  Thus we evaluate
$\Omega_\chi h^{2}$ and we determine $\rho_{\chi}$ by adopting a
standard rescaling procedure
\cite{Gaisser:1986ha}:

\be
\begin{split}
\rho_\chi&=\rho_0\,,
\qquad\qquad\qquad~~~ {\rm when}~~~ 
\Omega_\chi h^{2} \geq (\Omega_{CDM} h^{2})_{\rm min} \cr
\rho_\chi&= \rho_0 \frac{\Omega_\chi h^{2}}{(\Omega_{CDM} h^{2})_{\rm min}}\,,
~~~{\rm when}~~~ \Omega_\chi h^{2} < (\Omega_{CDM} h^{2})_{\rm min}
\end{split}
\label{eq:rescaling}
\ee

\noindent
Here $(\Omega_{CDM} h^{2})_{\rm min}$ is set to the 
value 0.098, as derived at 2$\sigma$ level from the results of
Ref.\,\cite{Dunkley:2008ie}. The neutralino relic abundance is evaluated 
with the procedure discussed in Ref.\,\cite{Bottino:2002ry,*Bottino:2003iu}. 

As for the velocity distribution of relic neutralinos in the
galactic halo we use here, for definiteness, the standard
isothermal distribution parametrized in terms of the local rotational velocity 
$v_{0}$. It is however to be recalled that
the actual distribution function could deviate sizably from
the isothermal one (see, for instance, Ref.\,\cite{Belli:2002yt} for
a systematic analysis of different categories of
distribution functions) or even depend on non-thermalized
effects \cite{Freese:2003na,*Bernabei:2006ya,*Duffy:2008dk}. 
Also the possible presence of a thick disk of dark matter could play a relevant role
in the capture of dark matter by celestial bodies \cite{Bruch:2009rp}. 
The local rotational velocity $v_{0}$ is set at three different representative 
values: the central value $v_{0} = 220$ Km s$^{-1}$ and two extreme values 
$v_{0} = 170$ Km s$^{-1}$ and $v_{0} = 270$ Km s$^{-1}$ which bracket the 
$v_{0}$ physical range. Associated to each value of $v_{0}$ we take a value 
of $\rho_{0}$ within its physical range established according to the procedure 
described in Ref.\,\cite{Belli:2002yt}. 
In conclusion, we will provide the numerical 
results of our analysis for the following three sets of astrophysical parameters: 
1) $v_{0} = 170$ Km s$^{-1}$, $\rho_{0} = 0.20$ GeV cm$^{-3}$; 
2) $v_{0} = 220$ Km s$^{-1}$, $\rho_{0} = 0.34$ GeV cm$^{-3}$; 
3) $v_{0} = 270$ Km s$^{-1}$, $\rho_{0} = 0.62$ GeV cm$^{-3}$. 
Note that these values of $\rho_{0}$ correspond to the case of maximal amount of non 
halo components to DM in the galaxy \cite{Belli:2002yt}.

The annihilation rate $\Gamma_{\rm ANN}$ is expressed in terms of
the capture rate by the formula \cite{Griest:1986yu}

\be
\Gamma_{\rm ANN}=\frac{C}{2} {\rm tanh}^{2} \left(\frac{t}{\tau_A}\right)\,,
\label{eq:Annihil}
\ee

\noindent
where $t$ is the age of the macroscopic body ($t= 4.5~{\rm Gyr}$ for
Sun and Earth), $\tau_A = (C C_A)^{-1/2}$, and $C_A$ is the
annihilation rate per effective volume of the body, given by

\be
C_A=\frac{<\sigma v>}{V_0} \left(\frac{m_\chi}{20~{\rm GeV}}\right)^{3/2}\,.
\ee

\noindent 
In the above expression $V_0$ is defined as 
$V_0=(3 m^{2}_{Pl} T / (2 \rho \times 10~{\rm GeV}))^{3/2}$
where $T$ and $\rho$ are the central temperature and the central
density of the celestial body. For the Earth ($T=6000 ~{\rm K}$,
$\rho= 13~{\rm g} \cdot {\rm cm}^{-3}$) $V_0= 2.3 \times 10^{25} {\rm
cm}^{3}$, for the Sun ($T=1.4 \times 10^{7}~{\rm K}$, $\rho= 150~{\rm g}
\cdot {\rm cm}^{-3}$) $V_0= 6.6 \times 10^{28}~{\rm cm}^{3}$. $\sigma$
is the neutralino--neutralino annihilation cross section and $v$ is
the relative velocity. The thermal average $<\sigma v>$ is calculated with all the
contributions at the tree level as in Ref.\cite{Bottino:1993zx}, 
with the further inclusion here of the two gluon annihilation final 
state \cite{Drees:1993bh}.

We recall that, according to Eq.\,\eqref{eq:Annihil}, in a given
macroscopic body the equilibrium between capture and annihilation
({\it i.e.} $\Gamma_{\rm ANN} \sim C/2$ ) is established only when $t \gtrsim
\tau_A$.  It is worth noticing that the neutralino density
$\rho_{\chi}$, evaluated according to Eq.\,\eqref{eq:rescaling}, enters 
not only in $C$ but also in $\tau_A$ (through $C$). Therefore the use of 
a correct value for $\rho_{\chi}$ (rescaled according to Eq.\,
\eqref{eq:rescaling}, when necessary)
is important also in determining whether or not the equilibrium is
already set in a macroscopic body.

Explicit calculations over the whole parameter space, given in the next 
sections, show that, whereas
for the Earth the equilibrium condition depends sensitively on the
values of the model parameters, in the case of the Sun equilibrium
between capture and annihilation is typically reached for the whole range of
$m_\chi$, due to the much more efficient capture rate implied by the
stronger gravitational field
\cite{Gould:1987ir,*Gould:1987ww,*Gould:1991ApJ}. 

The annihilation rate given above refers to a macroscopic body as a
whole. This is certainly enough for the Sun which appears to us as a
point source. On the contrary, in the case of the Earth, one has also
to define an annihilation rate referred to a unit volume at point $
\vec r$ from the Earth center:

\be
\Gamma_{\rm ANN} (r)= \frac{1}{2} <\sigma v> n^{2}(r)\,,
\ee

\noindent
where $n(r)$ is the neutralino spatial density which may be written as
\cite{Griest:1986yu}

\be
n(r) = n_0\,{\rm e}^{-{\tilde \alpha}\,m_{\chi}\,r^{2}}\,;
\label{eq:DMdistr}
\ee

\noindent
here $\tilde{\alpha} = 2 \pi G \rho / (3 T)$ and $n_0$ is such that

\be
\Gamma_{\rm ANN} = \frac{1}{2} <\sigma v> \int d^3r\, n^{2}(r)\,. 
\ee
\bigskip

\subsection{\label{sec:hadr}WIMP--nucleon cross section: hadronic uncertainties}

In Ref.\,\cite{Bottino:1999ei,*Bottino:2001dj} it is stressed that the couplings between
Higgs-bosons (or squarks) with nucleons, which typically play a crucial
role in the evaluation of the neutralino-nucleus cross section, suffer of large
uncertainties \cite{Ellis:2000ds,*Accomando:1999eg,*Corsetti:2000yq,*Feng:2000gh,*Ellis:2008hf,
*Giedt:2009mr}. 
Actually, these couplings are conveniently expressed in
terms of three hadronic quantities, {\it i.e.}:
\noindent
the pion--nucleon sigma term
\be
\sigma_{\pi N} = \frac{1}{2} (m_u + m_d) <N|\bar uu + \bar dd|N>\,,
\ee

\noindent the quantity $\sigma_{0}$, related to the size of the SU(3)
symmetry breaking,
\be
\sigma_{0}\equiv \frac{1}{2}(m_u+m_d) <N|\bar uu+\bar dd-2\bar ss|N>\,,
\ee
\noindent and the mass ratio $r=2 m_s/(m_u+m_d)$.

Because of a number of intrinsic theoretical and experimental
problems, the determination of these hadronic quantities is rather
poor. Conservatively, their ranges can be summarized as follows (we
refer to Refs.\,\cite{Bottino:1999ei,*Bottino:2001dj,Bottino:2008mf} for
details):

\be
41 \; {\rm MeV} \lesssim \sigma_{\pi N}\lesssim 73 \; {\rm MeV}\,,
\label{eq:q}
\ee

\be
\sigma_{0}=30\div40 \;{\rm MeV}\,,
\label{eq:0}
\ee
and
\be
r = 29 \pm 7.
\label{eq:r}
\ee

In the present paper, in order to display the influence of the
uncertainties due to the hadronic quantities on the signals at
neutrino telescopes, we will report our results for three different
sets of values for the quantities $(\sigma_{\pi N}\,,\sigma_{0}\,, r)$ 
as shown in Tab.\,\ref{tab:values}.
\begin{table}
\begin{center}
\begin{tabular} {||p{3cm}||*{3}{c|}|}
\hline
& & & \\ 
\centering{hadronic set} & $\,$ $\sigma_{\pi N}$[MeV] $\,$ &
$\,$ $\sigma_{0}$[MeV] $\,$ & $\quad r\quad$ \\ 
& & & \\
\hline
\centering{MIN}  &  41  &  40  &  25  \\
\hline
\centering{REF}  &  45  &  30  &  29  \\
\hline
\centering{MAX}  &  73  &  30  &  25  \\
\hline
\end{tabular}
\end{center}
\caption{Set of values for the hadronic quantities considered in the numerical analysis.}
\label{tab:values}
\end{table}
The set REF corresponds to the set of value referred to as {\it
reference point} in Ref.\,\cite{Bottino:2008mf}. The sets MIN and MAX 
listed in Tab.\,\ref{tab:values} 
bracket the range of hadronic uncertainties. 

In the case where the neutralino-nucleus interaction is dominated
by the exchange of Higgs bosons, it is straightforward
to estimate by how much the
capture rate $C$ is affected by the hadronic uncertainties.
Indeed, in this case the dominant term in the interaction amplitude
of the neutralino-nucleus scattering is provided by coupling between
the two CP-even Higgs bosons and the down-type quarks:
\be
g_d = \frac{2}{27}\left( m_{N} + \frac{23}{4}\,\sigma_{\pi N} + 
\frac{23}{5}\,r\,\left(\sigma_{\pi N} - \sigma_{0}\right) \right)\,,
\ee
where $m_N$ is the nucleon mass. Then:
\be
C_{\rm {MIN}}/C_{\rm {REF}} \simeq ( g_{d,\,{\rm MIN}}/g_{d,\,{\rm REF}} )^{2}\,,
\qquad
C_{\rm {MAX}}/C_{\rm {REF}} \simeq ( g_{d,\,{\rm MAX}}/g_{d,\,{\rm REF}} )^{2}\,.
\label{eq:ratioC}
\ee
Using the values of Table\,\ref{tab:values} for the
three sets of hadronic quantities, one finds
for $g_d$:
$g_{d,\,{\rm MIN}}$ = 99 MeV, $g_{d,\,{\rm REF}}$ = 290 MeV, 
$g_{d,\,{\rm MAX}}$ = 598 MeV, respectively. 
We thus conclude that, because of the hadronic uncertainties, the
capture rate in the case of set MIN is reduced by a factor $\sim9$ as compared
to the capture rate evaluated with the set REF, whereas $C$, evaluated with
set MAX, is enhanced by a factor $\sim4$.

The consequences over the annihilation rate $\Gamma_{\rm ANN}$ is more 
involved, since the capture rate $C$ enters in $\Gamma_{\rm ANN}$ not only 
linearly but also through $\tau_A$. When in the celestial body capture and annihilation
are in equilibrium ($t \gtrsim \tau_A$), one has $\Gamma_{\rm ANN} \sim C/2$; then
$\Gamma_{\rm ANN}$, as a function of the hadronic quantities, rescales as $C$
(see Eq.\,\eqref{eq:ratioC}); however, when the equilibrium is not realized, the 
uncertainties in
$\Gamma_{\rm ANN}$ can be much more pronounced. For instance, for
$t \ll \tau_A$, $\Gamma_{\rm ANN}$ is proportional to $C^2$, thus the rescaling
factors for $\Gamma_{\rm ANN}$ are the squares of those in Eq.\,\eqref{eq:ratioC}.

These estimates will be confirmed by the numerical analysis displayed
in the following section.

\subsection{Numerical evaluations}

\begin{figure}
\begin{center}
\begin{tabular}{cc}
\includegraphics[width=7.5cm,height=6.5cm]{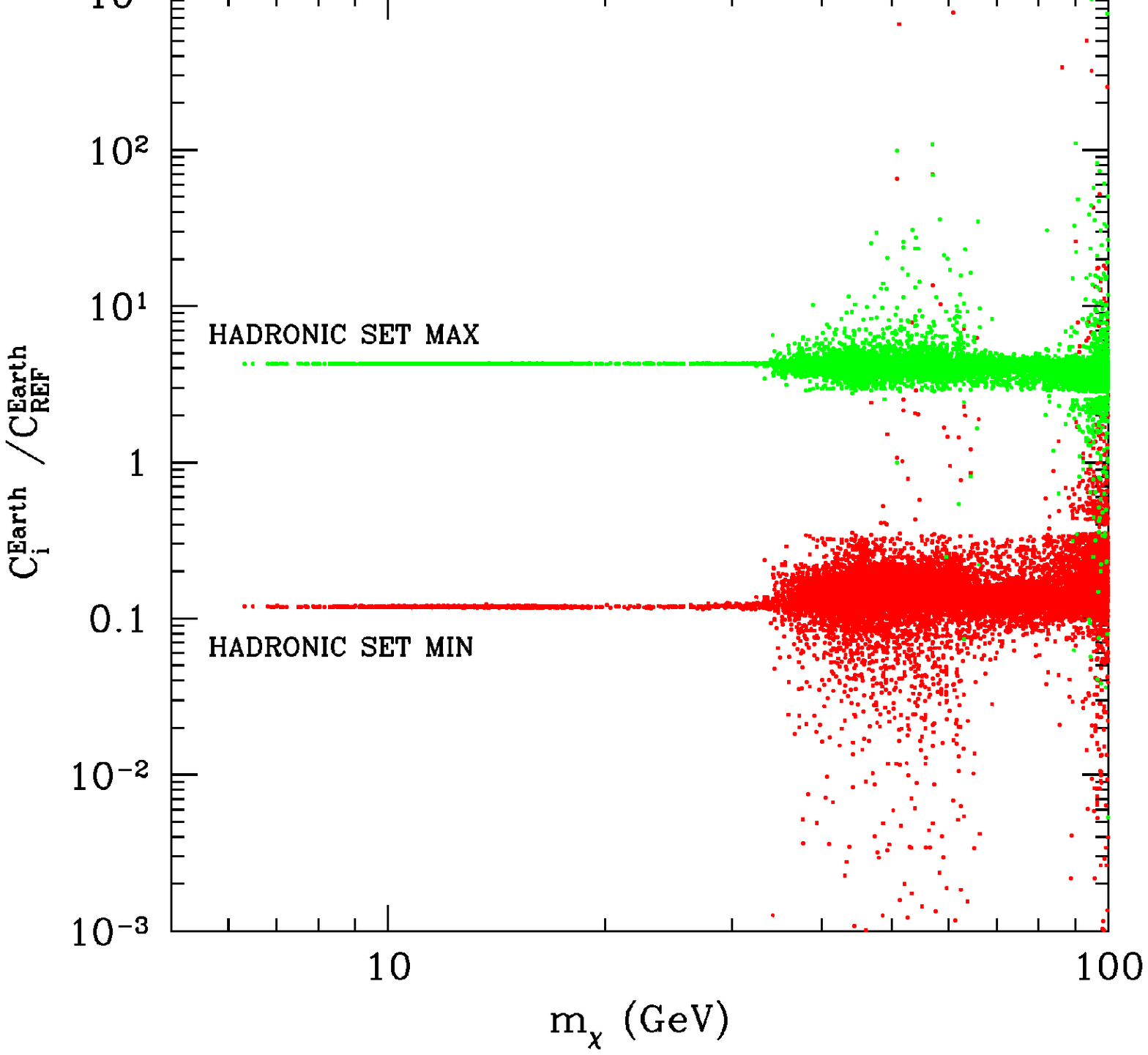}&
\includegraphics[width=7.5cm,height=6.5cm]{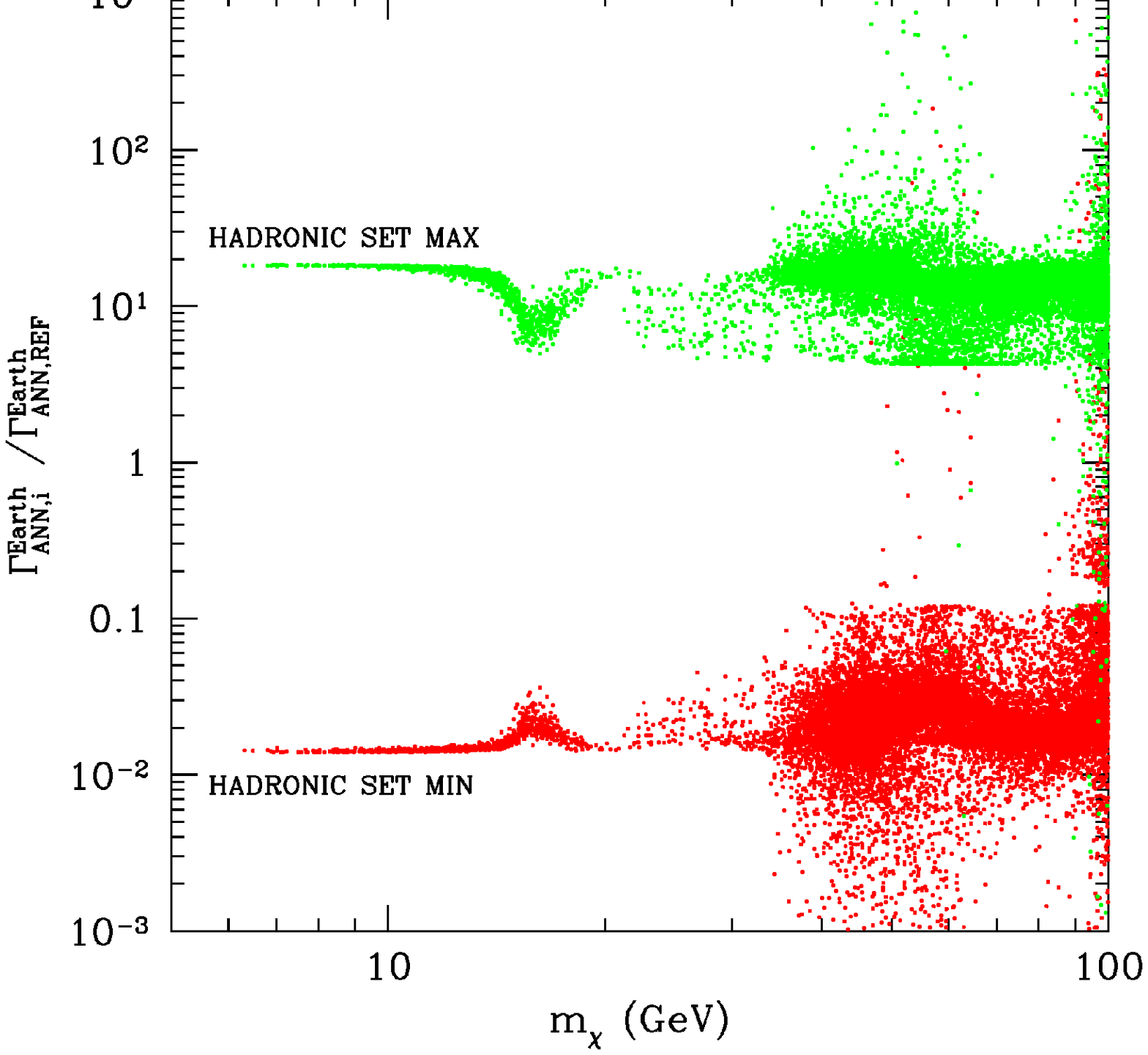}
\end{tabular}
\end{center}
\caption{\label{fig:CaptAnn_Earth}Ratios of capture rates (left panel) and 
annihilation rates (right panel), in the case of the Earth, calculated for the hadronic 
sets MIN and MAX with respect to the hadronic set REF. The local 
rotational velocity is set to its central value: $v_{0} = 220$ Km s$^{-1}$ 
($\rho_{0} = 0.34$ GeV cm$^{-3}$).}
\end{figure}

The left panel of Fig.\,\ref{fig:CaptAnn_Earth} shows the scatter plots for the 
ratios of the capture rates $C^{\rm Earth}_i/C^{\rm Earth}_{\rm REF}$
(where i = set MIN, set MAX), obtained with the scanning of the parameter 
space illustrated in Sect.\,\ref{sec:model}. One sees that, as anticipated in the previous 
section, the numerical values accumulate (most significantly for light masses),
around the numerical factors shown in Eq.\,\eqref{eq:ratioC}.

\begin{figure}
\begin{center}
\includegraphics[width=15cm,height=5cm]{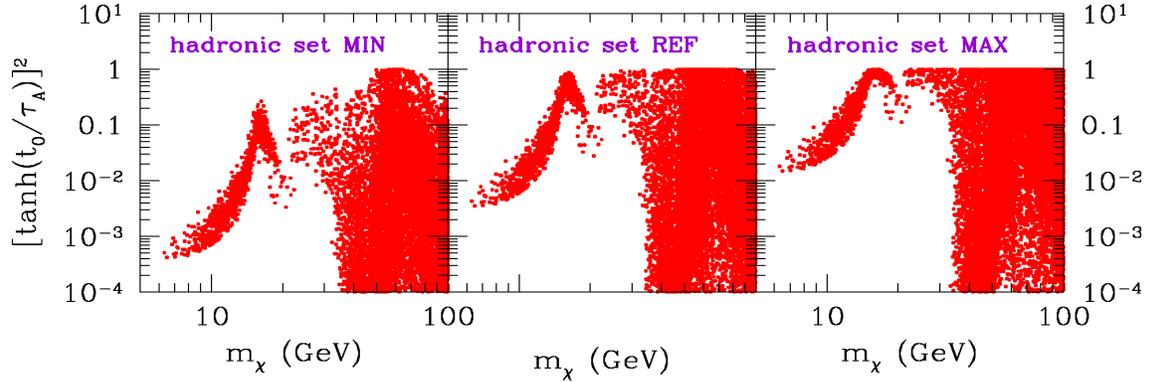}
\end{center}
\caption{\label{fig:eqFactor}Equilibrium factor $\tanh^2(t_0/{\tau}_A)$, in the 
case of the Earth, displayed for the hadronic sets MIN, REF and MAX. The local 
rotational velocity is set to its central value: $v_{0} = 220$ Km s$^{-1}$ 
($\rho_{0} = 0.34$ GeV cm$^{-3}$).}
\end{figure}

The scatter plots for the ratios
$\Gamma^{\rm Earth}_{{\rm ANN},i}/\Gamma^{\rm Earth}_{\rm ANN, REF}$ 
(where $i$ = set MIN, set MAX)
are displayed in the right panel of Fig.\,\ref{fig:CaptAnn_Earth}. 
As expected and discussed before, these numerical values
are much larger as compared to those of the ratios 
$C^{\rm Earth}_i/C^{\rm Earth}_{\rm REF}$,
since many supersymmetric configurations are not able to provide a
capture-annihilation equilibrium inside the Earth. This interpretation is 
validated by the scatter plots of Fig.\,\ref{fig:eqFactor}, where one sees 
that for most configurations the equilibrium factor $\tanh^2(t_0/{\tau}_A)$ 
deviate largely from the equilibrium value
$\tanh^2(t_0/{\tau}_A)$ = 1.

\begin{figure}
\begin{center}
\includegraphics[width=7.5cm,height=6.5cm]{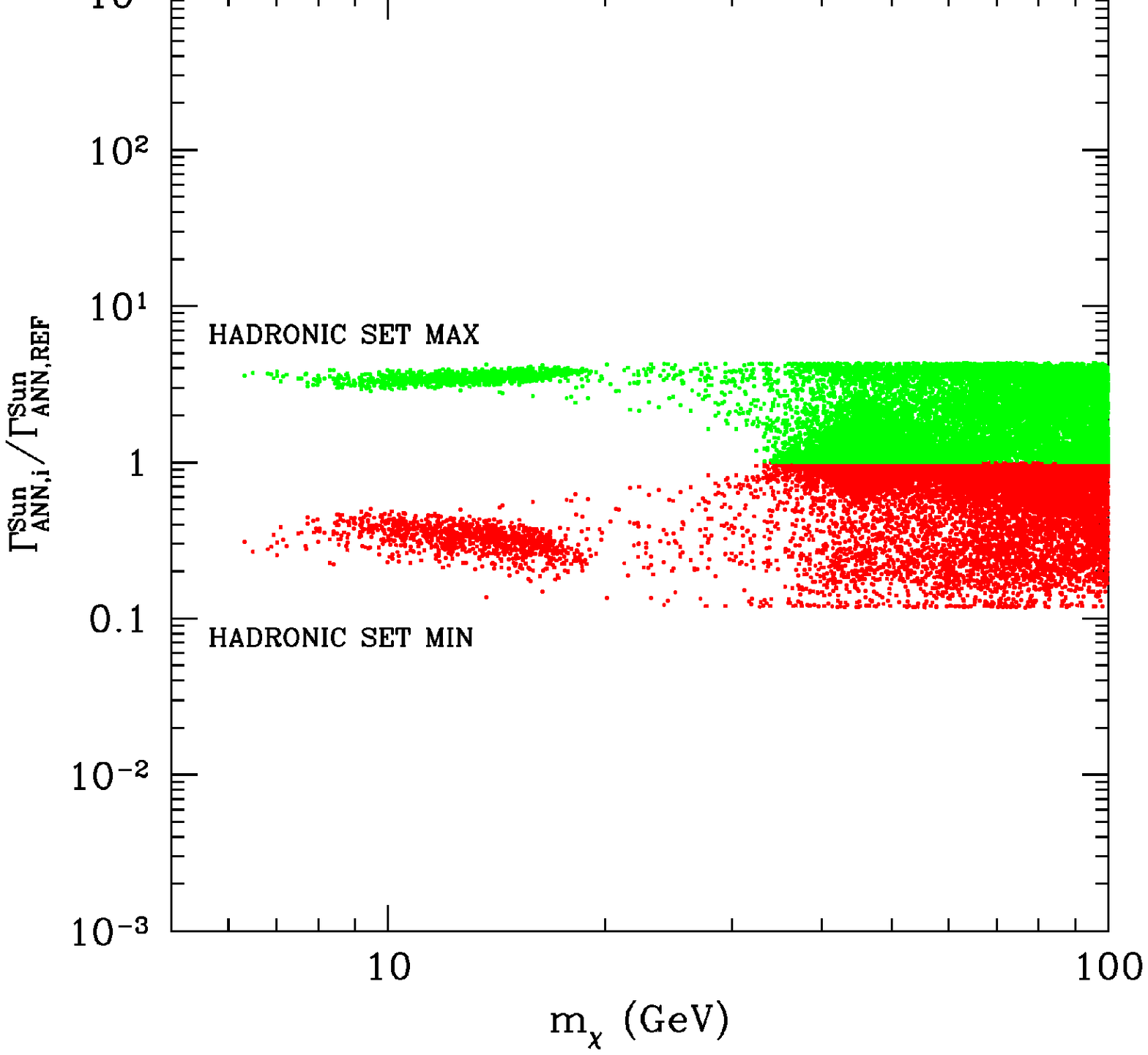}
\end{center}
\caption{\label{fig:gammaSun}Ratios of annihilation rates, 
in the case of the Sun, calculated for the hadronic 
sets MIN and MAX with respect to the hadronic set REF. The local 
rotational velocity is set to its central value: $v_{0} = 220$ Km s$^{-1}$ 
($\rho_{0} = 0.34$ GeV cm$^{-3}$).}
\end{figure}

The dependence of the annihilation rate for the Sun, 
$\Gamma^{\rm Sun}_{\rm ANN}$ on the 
hadronic uncertainties is shown in Fig.\,\ref{fig:gammaSun}. 
Since the capture-annihilation equilibrium is realized in the Sun
for all supersymmetric configurations of our model, one has here that
$\Gamma^{\rm Sun}_{{\rm ANN},i}/\Gamma^{\rm Sun}_{\rm ANN,REF} =
C^{\rm Sun}_{i}/C^{\rm Sun}_{\rm REF}$, which implies that
$\Gamma^{\rm Sun}_{\rm ANN,MAX}/\Gamma^{\rm Sun}_{\rm ANN,REF} \lesssim 4$ and
$\Gamma^{\rm Sun}_{\rm ANN,MIN}/\Gamma^{\rm Sun}_{\rm ANN,REF} \gtrsim 1/9$.
This is at variance with the case of the Earth which we have commented before.

Moreover, one notices from Fig.\,\ref{fig:gammaSun} 
that for many supersymmetric configurations
$\Gamma^{\rm Sun}_{\rm ANN}$ depends very slightly (or negligibly) on the
variations in the
the hadronic quantities. This is due to the fact that on many instances
the capture of neutralinos from the Sun is dominated by spin-dependent
cross-sections.

\section{\label{sec:neutrino}Calculation of neutrino fluxes}

\subsection{\label{sec:product}Neutrino production}

Once the Dark Matter particles are accumulated in the center of 
the Sun and the Earth, they can annihilate into leptons, 
quarks, gauge and higgs bosons. These particles then 
decay producing neutrinos. To calculate the neutrino 
spectra coming from each annihilation channel, lepton decay and 
quark hadronization process have to be considered.
In \cite{Cirelli:2005gh}, the authors have used a PYTHIA Monte Carlo 
simulation to calculate the spectra of neutrinos, coming from Dark Matter 
annihilation in the Sun and in the Earth, for the following channels: 
$b \bar{b},\,\tau \bar{\tau},\,c \bar{c},\,q \bar{q},\,g g$ (with $q$ = $u,\,d,\,s$ quarks). 
Three main differences and improvements have been implemented in 
\cite{Cirelli:2005gh} with respect to previous calculations. The first 
one is the prediction of the neutrino spectra for the different 
neutrino flavors: $\nu_{e}$, $\nu_{\mu}$ and $\nu_{\tau}$ (not only $\nu_{\mu}$, 
like in previous works). 
The second main improvement consists in an appropriate
implementation of the energy loss that hadrons and leptons can
undergo before decaying. Finally, the third difference is represented 
by the calculation of the neutrino spectra for 
light quarks $u,\,d,\,s$, that were usually neglected in previous calculations. 

We use the initial neutrino spectra reported in \cite{Cirelli:2005gh} for 
$b \bar{b},\,\tau \bar{\tau},\,c \bar{c},\,
q \bar{q}$ and $g g$ annihilation channels. 
The annihilation of two neutralinos can also produce two higgs
bosons or one gauge and one higgs boson in the final state, 
although these two channels (as well as the annihilation channels into $t \bar{t}$ 
and into two gauge bosons) are absent for $m_{\chi} \leq 80$ GeV 
(we remind that our model has an absolute lower limit on the higgs mass 
of 90 GeV).

\subsection{\label{sec:propagation}Neutrino propagation}

For a precise estimate of the neutrino flux at the detector site, it
is important to take into account the main processes that can occur
during the neutrino propagation: the oscillation and the incoherent
interaction with matter. These effects have been vastly analyzed in
\cite{Cirelli:2005gh,Blennow:2007tw}, and then applied to specific
model-dependent studies, see e.g.\,\cite{Barger:2007xf,*Liu:2008kz}.

The equations that describe the evolution of the neutrino spectra 
can be formally written using the density matrix formalism:
\be
\frac{d\rho}{dr}=-i\left[H,\rho\right] + \left.\frac{d\rho}{dr}\right|_{NC}
+ \left.\frac{d\rho}{dr}\right|_{CC}\,.
\label{eq:density}
\ee
The first term describes the oscillation of neutrinos in matter, where
the Hamiltonian is given by the sum of the mass matrix in the weak
basis $(\nu_{e}, \nu_{\mu}, \nu_{\tau})$ and the Wolfenstein
potential:
\be
H_{\textrm{w}} = \frac{M_{\textrm{w}}}{2E} \pm
\sqrt{2}\,G_{F}\,N_{e}\,\textrm{diag}(1,0,0)\,
\ee
holds for (anti-)neutrinos for the minus (plus) sign. We fix the
values for the neutrino mixing angles and squared mass differences to
the best fit values reported in \cite{Schwetz:2008er}, and we set the
mixing angle $\theta_{13}$ to zero. A different choice of
$\theta_{13}$ would marginally affect the prediction on the neutrino
flux, as reported in \cite{Cirelli:2005gh,Blennow:2007tw}.

The second term in Eq.\,\eqref{eq:density} takes into account the
neutrino energy loss and the reinjection due to neutral current
interactions. The last term, instead, represents the neutrino
absorption and $\nu_{\tau}$ regeneration through charge current
interactions.  For the explicit form of each term and for exhaustive
explanations we refer to \cite{Cirelli:2005gh}.

For Dark Matter annihilation inside the Sun, the integro-differential
equation for the density matrix has to be solved numerically to find
the neutrino spectra at the detector site. For simplicity, we neglect
the $\nu_{\tau}$ regeneration effect, since it provides only a
negligible correction for the WIMPs mass range of our interest:
$m_{\chi} \le 80$ GeV.

In the case of annihilations inside the Earth's core, the calculation
of the neutrino spectra can be further simplified.  Indeed, the
interactions with matter can be neglected, since the mean free paths
of neutrinos, in the core and in the mantle, are much bigger than the
Earth's radius $R_{\oplus}$ for $E_{\nu}\lesssim10$ TeV (for
anti-neutrinos the mean free paths are almost a factor two greater,
due to the difference in the cross-sections):
\be
\lambda^{core}=\frac{1}{\sigma_{\nu} N^{core}_{e}}\simeq3.6\times10^{4}
\frac{R_{\oplus}}{(E_{\nu}/{\text{GeV}})}\,,
\qquad
\lambda^{mantle}=\frac{1}{\sigma_{\nu} N^{mantle}_{e}}\simeq9.2\times10^{4}
\frac{R_{\oplus}}{(E_{\nu}/{\text{GeV}})}\,.
\ee
Therefore, for the propagation inside the Earth, it can be safely
taken into account only the oscillation effect.  Moreover, for
$E_{\nu}\gtrsim 1$ GeV, the dependence on the ``solar'' parameters
$\Delta m^{2}_{21}$ and $\theta_{12}$ is extremely weak and can be
neglected. Since we are considering vanishing $\theta_{13}$, Earth's
matter effects are negligible and the neutrino oscillation is
driven by the ``atmospheric'' parameters $\Delta m^{2}_{31}$ and
$\theta_{23}$. In this case, the main oscillation channel is the
$\nu_{\mu} \leftrightarrow \nu_{\tau}$ and the value of the
oscillation and the survival probability $P_{\alpha \beta}$ is simply
given by the vacuum two-flavors formula: 
\be P_{\alpha\beta}(r,E_{\nu})=\delta_{\alpha \beta}-\epsilon_{\alpha \beta}
\sin^{2}(2\theta)
\sin^{2}\left(1.27\,\frac{(\Delta m^{2}_{31}/\text{eV}^{2})(r/\text{km})}
{(E_{\nu}/\text{GeV})}\right)\,,
\ee
where the parameter $\epsilon_{\alpha \beta}$ is equal to 1 (-1) for
$\alpha=\beta\,(\alpha\neq\beta)$.

In the case of the Earth, the differential
muon-neutrino flux at the detector, as a function of the zenith angle
$\theta_{z}$, can be written as:
\be
\frac{dN_{\nu_{\mu}}}{dE_{\nu}\,d\cos \theta_{z}} =
\frac{\Gamma_{\rm ANN}}{4\pi R_{\oplus}^{2}}\,\sum_{f}\,BR_{f}\,
\left(G_{\mu\mu}(\theta_{z},E_{\nu})\,\frac{dN^{\nu_{\mu}}_{f}}{dE_{\nu}} + G_{\mu\tau}(\theta_{z},E_{\nu})\,\frac{dN^{\nu_{\tau}}_{f}}{dE_{\nu}}\right)\,,
\label{nuEarthDet}
\ee
where the function $G_{\alpha\beta}(\theta_{z},E_{\nu})$ encodes the
dependence on the oscillation probability and on the WIMP distribution 
inside the Earth. Using Eq.\,\eqref{eq:DMdistr}, we find the
following expression:
\be
G_{\alpha \beta}(\theta_{z},E_{\nu})=\frac{2\,(2\,m_{\chi}\,{\tilde \beta})^{3/2}}
{\pi^{1/2}\,R_{\oplus}}\,
\int^{y}_{0}\,dr\,
\exp \left[-2\,m_{\chi}\,{\tilde \alpha}\,\left(r^{2}+R^{2}_{\oplus}-r y\right)\right]
\,P_{\alpha \beta}(r,E_{\nu})\,,
\ee
with $y\equiv2\,R_{\oplus}\,\cos\theta_{n}$, $\theta_{n}\equiv
\pi-\theta_{z}$ and ${\tilde \beta}={\tilde \alpha}\,R^{2}_{\oplus}$. 
The differential muon anti-neutrino flux at the detector can be obtained by a formula
analogous to Eq.\,\eqref{nuEarthDet}.

\section{\label{sec:muon}Muon fluxes}

For the calculation of up-going muons we follow the formalism
described in 
\cite{1991crpp.bookG,*Gaisser:1984mx,*Gaisser:1985cm,*Fornengo:1999py}, 
to which we refer for specific details. In these references, it has 
been shown that the differential muon flux is given by the following 
expression:
\be
\frac{dN_{\mu}}{d\cos\theta_{z} dE_{\mu}}=
N_{A}~\frac{1}{a + b E_{\mu}}\int^{\infty}_{E_{\mu}}dE_{\nu}~
\int^{E_{\nu}}_{E_{\mu}}
dE^{\prime}_{\mu}~\frac{dN_{\nu_{\mu}}}{d\cos\theta_{z}dE_{\nu}}~
\frac{d\sigma_{\nu_{\mu}}(E_{\nu},E^{\prime}_{\mu})}{dE^{\prime}_{\mu}}\,.
\label{eq:diffFlux}
\ee
The quantities $a$ and $b$ parametrize the energy loss due to
ionization and to radiative effects, $N_{A}$ is the Avogadro's number and 
$d\sigma_{\nu_{\mu}}/dE^{\prime}_{\mu}$ is the differential 
charge current cross section, which is mainly due to deep inelastic 
scattering, for energies $E_{\nu}>1$ GeV. 
The total muon flux is then divided in through-going and
stopping muons:
\be
\Phi^{S,T}_{\mu}(\cos \theta_{z})=\frac{1}{A(L_{min},\theta_{z})} \int^{\infty}_{E^{th}_{\mu}}
d E_{\mu}\,\frac{d N_{\mu}}{d \cos\theta_{z} d
E_{\mu}}\,A^{S,T}(L(E_{\mu}),\theta_{z})\,,
\label{eq:finalmuons}
\ee
with $E^{th}_{\mu}$ being the energy threshold of the detector for
upward-going muons, $L$ the muon range in water (
$L=\frac{1}{b}\ln\frac{a + b\,E_{\mu}}{a + b\,m_{\mu}}$
) and $L_{min}\equiv
L(E^{th}_{\mu})$. The function $A^{S,T}(L,\theta_{z})$ represents the
effective area for through-going and stopping muons, while,
$A(L_{min},\theta_{z})$ is the total effective area of the detector,
i.e. the projected area that corresponds to internal path-lengths
longer than $L_{min}$, for a fixed value of the zenith angle
$\theta_{z}$.

The classification of upward-going muons into the two subcategories
reported above is strictly detector-dependent, since it depends on
the shape and the size of the detector. For a detector with cylindrical
geometry (with radius $R$ and height $H$), it has been shown in
\cite{Lipari:1998rf} that the function $A(L,\theta_{z})$ acquires the
form:
\be
A(L,\theta_{z})=2 R H \sin \theta_{z} \sqrt{1-x^{2}}\,+\,2 R^{2} |\cos
\theta_{z}| \left[ \cos^{-1} x -3 x \sqrt{1-x^{2}} \right]
\Theta(L_{max}(\theta_{z})-L)\,,
\label{eq:effarea}
\ee
with $x=L \sin \theta_{z}/2 R$ and $L_{max}(\theta_{z})=
\textrm{min}\left[2 R/ \sin \theta_{z},
H/|\cos \theta_{z}|\right]$. The effective area for stopping muons
$A^{S}(L,\theta_{z})$ is given by Eq.\,\eqref{eq:effarea} and the one
for through-going is $A^{T}(L,\theta_{z})\equiv \left[
A(L_{min},\theta_{z})-A(L,\theta_{z}) \right] $.

We will focus our analysis to the SK detector. Indeed, even if the
recent results of Ice-Cube 22-strings \cite{Abbasi:2009uz} improve the
SK bound in the high mass region ($m_{\chi}\gtrsim$ 200 GeV for the
hard channel and $m_{\chi}\gtrsim$ 500 GeV for the soft channel), they
do not constrain the parameter space at low masses.  The Ice-Cube
detector, provided with the Deep-Core arrays \cite{Resconi:2008fe},
will improve significantilly the SK bound for mass $m_{\chi}\gtrsim$
40 GeV, as has been shown in \cite{Wikstrom:2009kw}, but not for the
low mass range in which we are mainly interested: 7 GeV 
$\leq m_{\chi} \leq$ 80 GeV.

The SK detector has $R=16.9$ m, $H=36.2$ m and an energy threshold
$E^{th}_{\mu}=1.6$ GeV (that corresponds to $L_{min}=7$ m). Using
these features the effective area can be calculated through
Eq.\,\eqref{eq:effarea} and, with Eq.\,\eqref{eq:finalmuons}, the
muons coming from Dark Matter annihilation can be divided in
through-going and stopping muons. The same formalism can be applied to
the calculation of the expected muons background coming from
atmospheric neutrinos. We use the atmospheric neutrino flux from Honda
et al.\,\cite{Honda:2004yz} and we reproduce with great accuracy the
zenith angle distribution for stopping and through-going muons, as
predicted by the SK collaboration. For the muon energy loss in rock 
and in water, we use the tabulated values reported in \cite{Lohmann:1985qg}. 
Following the analysis for through-going muons of \cite{Desai:2004pq}, 
the SK limit on the muon flux can be defined as: 
\be
\Phi(\theta_{z};\,90\%\:\text{C.L.})=\frac{N_{90}}{A(L_{min},\theta_{z}) \times T}\,,
\label{eq:Limit90}
\ee
where $N_{90}$ is the upper Poissonian limit at the 90$\%$ C.L., given
the measured events and the muons background from atmospheric
neutrino, and $T$ is the detector lifetime. We do not consider the
detector efficiency, since it is almost equal to $100\%$ for
upward-going muons.
Using Eq.\,\eqref{eq:Limit90} and the SK data collected from May 1996 to July 2001 
\cite{Ashie:2005ik}, we find the following limits on through-going ($\Phi^{T}_{\mu}$) 
and stopping ($\Phi^{S}_{\mu}$) muons:
\be
\Phi^{T}_{\mu,Earth}\lesssim 0.8\times 10^{-14}\,\,{\rm cm^{-2}\,s^{-1}}
\quad \text{at 90}\% \text{C.L.}\,,
\label{eq:thru_earthlimit}
\ee
\be
\Phi^{S}_{\mu,Earth}\lesssim 0.5\times 10^{-14}\,\,{\rm cm^{-2}\,s^{-1}}
\quad \text{at 90}\% \text{C.L.}\,,
\label{eq:stop_earthlimit}
\ee
\be
\Phi^{T}_{\mu,Sun}\,\,\,\,\lesssim 1.2\times 10^{-14}\,\,{\rm cm^{-2}\,s^{-1}}
\quad \text{at 90}\% \text{ C.L.}\,,
\label{eq:thru_sunlimit}
\ee
\be
\Phi^{S}_{\mu,Sun}\,\,\,\,\lesssim 0.5\times 10^{-14}\,\,{\rm cm^{-2}\,s^{-1}}
\quad \text{at 90}\% \text{ C.L.}\,,
\label{eq:stop_sunlimit}
\ee
The limits for the Earth are obtained considering the angular bin
$-1.0\le \cos\theta_{z}\le-0.9$, while the values reported for the Sun
are the average limits obtained varying the zenith angle from
$\cos\theta_{z}=-1.0$ to $\cos\theta_{z}=0$.

The values reported in Eqs.\,\eqref{eq:thru_earthlimit}$\div$\eqref{eq:stop_sunlimit} 
have to be compared with the muon flux induced by neutralinos annihilation. 
For the case of the Earth, we fix the angular opening to 
$-1.0\le \cos\theta_{z}\le-0.9$ also for the calculation of muons 
coming from Dark Matter. For the case of the Sun, we divide the muons 
in stopping and through-going using the SK effective area 
averaged over the zenith angle. 

In the calculation of the muon flux, we neglect the kinematical angle
between the neutrino and muon direction, which can be relatively large
for muons close to threshold. In any case, the average deflection angle
is at most of the same order of the angular bin over which we integrate
our signal, for the stopping and through going muons. Considering also
the detector resolution, to neglect the kinematical angle does not affect
our results in a relevant way. This is confirmed by the quite good agreement
we obtain in our calculation of the atmospheric--neutrino events with the
SK evaluation \cite{skthesis:2002}.

\section{\label{sec:results}Results and conclusions}

In this section we give our results for the muon fluxes expected at a
neutrino telescope with a threshold muon energy of 1.6 GeV, generated by
annihilation of light neutralino pair-annihilation in the Earth and in
the Sun.

\subsection{\label{sec:earth_generic}Fluxes from the Earth}

The upper panel of Fig.\,\ref{fig:Earth_thru} displays the scatter plots for 
the expected muon flux integrated over the muon energy for $E_{\mu} \geq$ 1.6 GeV 
for the upward through-going muons. The three columns refer to the evaluation of the
fluxes using in turn the three different set of hadronic quantities
defined in Sect.\,\ref{sec:hadr}.

The main features common to the three scatter plots are easily
interpretable in terms of the following properties:

i) The various peaks
for $m_{\chi} \lesssim$ 40 GeV are due to resonant capture of neutralinos
on oxygen, silicon and magnesium; indeed, these elements are almost as
abundant in Earth as iron, which is the most relevant target nucleus for
the capture of neutralinos of higher mass. The dip at  $m_{\chi} \sim$ 45 GeV
is a consequence of a depletion of the neutralino local density, implied
by the rescaling recipe of Eq. (2) and a resonant effect in the (Z-exchange) neutralino pair
annihilation  when $m_{\chi} \lesssim m_Z/2$ (note that the neutralino
relic abundance is inversely proportional to the neutralino pair-annihilation).

ii) Also the fact that the muon signal for light neutralinos
($m_{\chi} \lesssim$ 25-30 GeV) is lower than the one at higher masses can be
understood. Indeed, for light neutralino masses the branching ratio of the
annihilation process into the $\tau - \bar{\tau}$ final state, which is the one
with the highest neutrino yield per annihilation, is suppressed. This last
property being in turn due to the fact that, for these masses, the final state
in  $b - \bar{b}$   in the annihilation cross section has to be the dominant
one in order to keep the neutralino relic abundance below its cosmological
upper bound \cite{Bottino:2002ry,*Bottino:2003iu}.

iii) Moreover, lower $m_{\chi}$ masses
imply softer neutrino spectra which entail fewer muons above threshold.

The comparison of the fluxes in the three columns shows how relevant can be
the role of the size of the hadronic quantities on the final outputs.
The suppression (enhancement) of the flux in the case of the set MIN (MAX) as
compared to the flux for the set REF are set by the numerical factors
previously discussed for $\Gamma^{\rm Earth}_{\rm ANN}$. This entails that,
whereas the overall muon flux is completely below the present experimental 
bound in the case of the minimal set of the hadronic quantities, some part 
of the spectrum would emerge sizably above the limit for neutralino masses 
$m_{\chi} \gtrsim$ 50 GeV for the other sets. In the case of set MAX, 
neutralino configurations with masses $m_{\chi} \sim$ 15 GeV or $m_{\chi} \sim$ 
25-30 GeV might produce some measurable signal.

Since also the dependence of the muon signals on the astrophysical parameters $v_0$
and $\rho_0$ is important, in the lower panel of Fig.\,\ref{fig:Earth_thru} 
we display the through-going fluxes for the three
representative values of $v_0$ and $\rho_0$ which we discussed in Sect.\,\ref{sec:capture}. 
The overall increase in the fluxes in moving from left to right is
essentially due to the increase in the value of the local DM density.
In these scatter plots the hadronic quantities are set to the value REF.

The fluxes for upward stopping-muons from the Earth are given in
Fig.\,\ref{fig:Earth_stop}. 
The scheme of this figure is the same as the one of the
previous Fig.\,\ref{fig:Earth_thru}: the dependence of the fluxes on the hadronic quantities
can be read in the upper panel, the one on the astrophysical parameters is displayed in 
the lower panel.

Because of the uncertainties affecting the evaluations of the muon fluxes, mainly
due to the hadronic quantities, we cannot convert these results in terms of absolute
constraints on supersymmetric configurations. However, we can conclude that
the analysis of stopping muons from the Earth can have an interesting discovery
potential not only  for masses above 50 GeV, but also for light neutralinos
with $m_{\chi} \sim$ 15 GeV or $m_{\chi} \sim$ 25-30 GeV. 
Notice however that the neutralino configurations which provide the highest 
values of the muon fluxes, mainly at  $m_{\chi} \sim$ 50-70 GeV, are actually 
disfavored by measurements of WIMP direct detection \cite{Directdama:09} which
have their maximal sensitivity in this mass range. 

\subsection{\label{sec:sun_generic}Fluxes from the Sun}

The fluxes of upward through-going muons and for stopping muons from the Sun
are provided in Fig.\,\ref{fig:Sun_thru} and in Fig.\,\ref{fig:Sun_stop}, respectively.
The schemes of these figures is the same as the ones of the
previous Fig.\,\ref{fig:Earth_thru} and Fig.\,\ref{fig:Earth_stop}, respectively: 
the dependence of the fluxes on the hadronic quantities
can be read in the upper panels of Fig.\,\ref{fig:Sun_thru} and Fig.\,\ref{fig:Sun_stop}, 
the one on the astrophysical parameters is
displayed in the lower panels of the same figures. 

From these results one notices that through-going muons can only
be relevant for neutralinos  with masses $m_{\chi} \gtrsim$ 50 GeV or
$m_{\chi} \sim$ 35-40 GeV, whereas stopping muons can potentially provide
information also on some supersymmetric configurations with masses down to
$m_{\chi} \sim$ 7 GeV, in the favorable cases of high values of the hadronic
quantities and of the astrophysical parameters.

\subsection{\label{sec:dama}Fluxes of stopping muons for supersymmetric
configurations selected by the DAMA/LIBRA annual modulation data}

Now we give the expected upward muon fluxes from Earth and Sun which
would be produced by neutralino configurations which fit the annual modulation
data of the DAMA/LIBRA experiment \cite{Bernabei:2008yi}. As before, for definiteness 
the analysis is performed in the framework of the isothermal sphere.
The selection of the supersymmetric configurations is performed on the 
basis of the analysis carried out in Ref.\,\cite{Bottino:2008mf}:
for any set of astrophysical parameters and hadronic quantities, from the whole
neutralino population are extracted the configurations which fit the experimental
annual modulation data, and the relevant muon fluxes are evaluated.
As for the yearly modulation data, we consider both outputs of the
experimental analysis of the DAMA Collaboration: those where the channelling
effect \cite{Bernabei:2007hw} is included as well as those where this effect is neglected.
We recall that the way by which the channeling effect has to be taken into
account in the analysis is still under study; thus the actual physical outputs
in the analysis of the experimental data in terms of specific DM candidates could stay
mid-way, between the case defined as channeling and the no-channeling one, respectively.

We only report the results for stopping muons, since, as we have seen above,
this is the category of events which can provide the most sizable signals.
To avoid proliferation of figures, only fluxes calculated with the set REF for
the hadronic quantities is reported here.

Fig.\,\ref{fig:DAMA_Earth_stop} displays the fluxes for the upward stopping muons expected 
from the Earth in case of no-channeling (upper panel) and in the case of channeling 
(lower panel). The corresponding fluxes from the Sun are shown in Fig.\,\ref{fig:DAMA_Sun_stop}.

We note that depending on the role of channeling in the extraction of the physical
supersymmetric configurations, the stopping muon fluxes can have a discovery potential
with an interesting complementarity between the signals from the two
celestial bodies: whereas the flux from the Earth cannot give insights into
neutralino masses below about 15 GeV, the flux from the Sun would potentially be able to
measure effects down to $m_{\chi} \sim$ 7 GeV.

It is worth remarking that under favorable conditions provided by the actual values of the
involved parameters, a combination of the annual modulation data and of measurements at
neutrino telescopes could help in pinning down the features of the DM particle and in
restraining the ranges of the many quantities (of astrophysical and particle-physics
origins) which enter in the evaluations and still suffer from large uncertainties.

We stress once more that the present analysis, for definiteness, was performed only in the
standard case of a halo DM distribution function given by an isothermal sphere. Use of 
different halo distributions such as those mentioned in Sect.\,\ref{sec:capture} 
could modify sizably the role of specific supersymmetric configurations. 

We wish here to recall that indirect signals of light neutralinos could 
also be provided by future measurements of cosmic antideuterons in space 
\cite{Donato:1999gy,*Baer:2005tw,*Donato:2008yx}. 
Finally, investigations at the Large Hadron Collider 
will hopefully provide a crucial test bench for the very existence of these 
light supersymmetric stable particles \cite{Bottino:2008xc,*lhc:09}.

\vspace{0.3cm}
\begin{center}
{\bf Acknowledgments}
\end{center}

We  acknowledge Research Grants funded jointly by Ministero dell'Istruzione,
dell'Universit\`a e della Ricerca, by Universit\`a di Torino and by
Istituto Nazionale di Fisica Nucleare within the {\sl Astroparticle Physics
Project}. S. S. acknowledges support of the WCU program (R32-2008-000-10155-0) of
the National Research Foundation of Korea.

\newcommand{\eprint}[1]{
arXiv: \href{http://arxiv.org/abs/#1}{\texttt{#1}}
}

\bibliographystyle{JHEP}
\bibliography{DMnu_paper}

\clearpage

\begin{figure}
\begin{center}
\begin{tabular}{c}
\includegraphics[width=15cm,height=5cm]{./Earth_thru_A1_220_3panels.eps}\\\\
\includegraphics[width=15cm,height=5cm]{./Earth_thru_reference_3panels.eps}
\end{tabular}
\end{center}
\caption{\label{fig:Earth_thru} Upward through-going muon flux, generated by 
light neutralino pair-annihilation inside the Earth.
The upper panel shows the dependence of the muon flux on the 
hadronic quantities, for fixed values of the astrophysical parameters: 
$v_{0} = 220$ Km s$^{-1}$ and $\rho_{0} = 0.34$ GeV cm$^{-3}$. 
The lower panel shows the dependence of the muon flux on the local rotational 
velocity $v_{0}$ and the total DM density $\rho_{0}$, for the hadronic set REF. 
The horizontal line represents the 
experimental limit on through-going muons from the Earth obtained using 
the SK data, see Eq.\,\eqref{eq:thru_earthlimit}.}
\end{figure}

\begin{figure}
\begin{center}
\begin{tabular}{c}
\includegraphics[width=15cm,height=5cm]{./Earth_stop_A1_220_3panels.eps}\\\\
\includegraphics[width=15cm,height=5cm]{./Earth_stop_reference_3panels.eps}
\end{tabular}
\end{center}
\caption{\label{fig:Earth_stop} The same as Fig.\,\ref{fig:Earth_thru}, but in the 
case of upward stopping muons. In this case, the horizontal line refers to the 
experimental limit on stopping muons from the Earth obtained using 
the SK data, see Eq.\,\eqref{eq:stop_earthlimit}.}
\end{figure}

\begin{figure}
\begin{center}
\begin{tabular}{c}
\includegraphics[width=15cm,height=5cm]{./Sun_thru_A1_220_3panels.eps}\\\\
\includegraphics[width=15cm,height=5cm]{./Sun_thru_reference_3panels.eps}
\end{tabular}
\end{center}
\caption{\label{fig:Sun_thru} The same as Fig.\,\ref{fig:Earth_thru}, 
but in the case of light neutralino pair-annihilation inside the Sun. 
In this case, the horizontal line refers to the 
experimental limit on through-going muons from the Sun obtained using 
the SK data, see Eq.\,\eqref{eq:thru_sunlimit}.}
\end{figure}

\begin{figure}
\begin{center}
\begin{tabular}{c}
\includegraphics[width=15cm,height=5cm]{./Sun_stop_A1_220_3panels.eps}\\\\
\includegraphics[width=15cm,height=5cm]{./Sun_stop_reference_3panels.eps}
\end{tabular}
\end{center}
\caption{\label{fig:Sun_stop} The same as Fig.\,\ref{fig:Earth_thru}, 
but in the case of light neutralino pair-annihilation inside the Sun and of upward 
stopping muons. In this case, the horizontal line refers to the 
experimental limit on stopping muons from the Sun obtained using 
the SK data, see Eq.\,\eqref{eq:stop_sunlimit}.}
\end{figure}

\begin{figure}
\begin{center}
\begin{tabular}{c}
\includegraphics[width=15cm,height=5cm]{./Earth_stop_dama_nochann_3panels.eps}\\\\
\includegraphics[width=15cm,height=5cm]{./Earth_stop_dama_chann_3panels.eps}
\end{tabular}
\end{center}
\caption{\label{fig:DAMA_Earth_stop}Upward stopping muon flux, generated by 
light neutralino pair-annihilation inside the Earth. The configurations 
displayed are only the ones compatible with 
the DAMA/LIBRA annual modulation region, obtained without including the 
channeling effect (upper panel) and including the channeling effect (lower panel). 
The three columns show the results for the different sets of astrophysical 
parameters, defined in Sect.\,\ref{sec:capture}. The set REF is used for 
the hadronic quantities. The horizontal line represents the 
experimental limit on stopping muons from the Earth obtained using 
the SK data, see Eq.\,\eqref{eq:stop_earthlimit}.}
\end{figure}

\begin{figure}
\begin{center}
\begin{tabular}{c}
\includegraphics[width=15cm,height=5cm]{./Sun_stop_dama_nochann_3panels.eps}\\\\
\includegraphics[width=15cm,height=5cm]{./Sun_stop_dama_chann_3panels.eps}
\end{tabular}
\end{center}
\caption{\label{fig:DAMA_Sun_stop}The same as Fig.\,\ref{fig:DAMA_Earth_stop}, 
but in the case of light neutralino pair-annihilation inside the Sun. 
In this case, the horizontal line refers to the 
experimental limit on stopping muons from the Sun obtained using 
the SK data, see Eq.\,\eqref{eq:stop_sunlimit}.}
\end{figure}


\clearpage

\end{document}